\documentclass[11pt,twoside]{article}
\usepackage[utf8]{inputenc}
\usepackage[T1]{fontenc}
\usepackage{amsmath}
\usepackage{amsfonts}
\usepackage{amssymb}
\usepackage{xcolor}
\usepackage{mathtools}
\usepackage{hyperref}
\usepackage{doi}
\usepackage{array}
\newcolumntype{s}{>{\footnotesize}l}
\usepackage{booktabs}
\hypersetup{pdfborder={0 0 0},
	colorlinks=true,
	linkcolor=black,
	citecolor=black,
	urlcolor=black}
\usepackage{cleveref}
\usepackage{fourier}

\date{Compiled \today}

\providecommand{\tightlist}{%
  \setlength{\itemsep}{0pt}\setlength{\parskip}{0pt}}
\usepackage{verbatim}
\usepackage{longtable}

\newlength{\cslhangindent}
\newlength{\csllabelwidth}
\newlength{\cslentryspacingunit} 
  {}%
  {\par}
\newenvironment{CSLReferences}[2] 
 {
  \setlength{\parindent}{0pt}
  \ifodd #1
  \let\oldpar\par
  \def\par{\hangindent=\cslhangindent\oldpar}
  \fi
  \setlength{\parskip}{#2\cslentryspacingunit}
 }%
 {}
\usepackage{calc}

\newcommand{\pkg}{\textbf}

\usepackage[top=3cm, bottom=3cm, left=2.25cm, right=2.25cm]{geometry}
\usepackage[calcwidth,  sf,  big, compact]{titlesec}
\titleformat{\section}
{\normalfont\scshape \Large}{\thesection}{1em}{}
\usepackage{fancyhdr}
\usepackage{setspace}
\usepackage[authoryear]{natbib}
\usepackage{cleveref}

\titleformat{\section}[block]{\large \bf }
{  {\thesection.}}{4pt}{   }
\titleformat{\subsection}[block]{\itshape}
{  {\thesubsection.}}{4pt}{   }

\title{An R Package for Modelling Excess Lifetimes}
\author{L\'eo R. Belzile\thanks{HEC Montréal, Department of Decision Sciences, HEC Montréal (\texttt{leo.belzile@hec.ca})}
}
\date{}

\begin{document}
\maketitle

\abstract{%
The \texttt{longevity} \textbf{R} package provides provide maximum likelihood estimation routine for modelling of survival data that are subject to non-informative censoring and truncation mechanisms. It includes a selection of 12 parametric models of varying complexity, with a focus on tools for extreme value analysis and more specifically univariate peaks over threshold modelling. The package comes with visual diagnostics that account for the sampling scheme for lifetime data, utilities for univariate threshold selection, nonparametric maximum likelihood estimation, goodness-of-fit diagnostics and model comparisons tools. The different methods therein are illustrated using aggregated tabular data of longevity from Japan, and truncated lifelengths Dutch records.
}

\hypertarget{introduction-and-motivation}{%
\section{Introduction and motivation}\label{introduction-and-motivation}}

Many datasets collected by demographers for the analysis of human longevity have unusual features that are not commonly encountered and for which limited software implementations exists. Data for the statistical analysis of (human) longevity can take the form of aggregated counts per age at death, or most commonly life trajectory of individuals. Due to the costs of data validation, only the largest records are typically available: this is also in line with extreme value theory, which dictates that only these events should be used for inference. Many databases provide birth and death dates of individuals who died above age 100 years or more. Lifetimes are often interval truncated (only age at death of individuals dying between two calendar dates are recorded) or left truncated and right censored (when data of individuals still alive at the end of the collection period are also included). Another frequent format is death counts, aggregated per age band. In all instances, censoring and truncation is of administrative nature and thus non-informative about death.

To showcase the functionalities of the package, we consider Dutch and Japanese lifelengths.
The \texttt{dutch} data base contains the age at death (in days) of Dutch who died above age 92 between 1986 and 2015; there data were obtained from Statistics Netherlands and analyzed in Einmahl, Einmahl, and Haan (2019) and Belzile et al. (2022). Records are interval truncated, as people are included in the database only if they died during the collection period aged 92 years old and above. The truncation bounds for each individual can be obtained by subtracting from the endpoints of the sampling frame the birth dates, with left truncation bound \(\texttt{ltrunc}=\min\{92 \text{ years}, 1986.01.01 - \texttt{bdate}\}\) and right truncation \(\texttt{rtrunc} = 2015.12.31- \texttt{bdate}\). In addition, there are 226 interval-censored and interval-truncated records for which month and year of birth and death are known, as opposed to exact dates.

The second data base analyzed therein consists of counts of Japanese centenarians by age band; the data in \texttt{japanese2} are reproduced from Table 10.3 of the monograph \emph{Exceptional lifespans} (Maier, Jeune, and Vaupel 2021): only individuals who died above age 100, rounded down to the nearest year are given. The data for female Japanese are reproduced in Table \ref{tab:tbl-japanese-women}. The data were constructed using the extinct cohort method and are stratified by both birth cohort and sex. Such data are interval censored and right truncated: more specifically for this dataset, the Japanese lifetimes are interval censored between \(\texttt{age}\) and \(\texttt{age} + 1\) and right truncated at the age reached by the oldest individuals of their birth cohort in 2020. The \texttt{count} variable lists the number of instances in the contingency table, and serves as a weight variable. Assuming that the ages at death are independent and identically distributed with distribution function \(F(\cdot; \boldsymbol{\theta})\), the log likelihood for exceedances \(y_i = \texttt{age}_i - u\) is
\begin{align*}
\ell(\boldsymbol{\theta}) = \sum_{i: \texttt{age}_i  > u}n_i \left[\log \{F(y_i+1; \boldsymbol{\theta}) - F(y_i; \boldsymbol{\theta})\} - \log F(r_i - u; \boldsymbol{\theta})\right]
\end{align*}
where \(n_i\) is the count of the number of individuals in cell \(i\) and \(r_i > \texttt{age}_i+1\) is the right truncation limit for that cell, i.e., the maximum age that could have been achieved for that birth cohort by the end of the data collection period.

\begin{table}

\caption{\label{tab:tbl-japanese-women}Death count by birth cohort for female Japanese.}
\centering
\begin{tabular}[t]{rrrrrrr}
\toprule
age & 1874-1878 & 1879-1883 & 1884-1888 & 1889-1893 & 1894-1898 & 1899-1900\\
\midrule
100 & 1648 & 2513 & 4413 & 8079 & 16036 & 9858\\
101 & 975 & 1596 & 2921 & 5376 & 11047 & 7091\\
102 & 597 & 987 & 1864 & 3446 & 7487 & 4869\\
103 & 345 & 597 & 1153 & 2230 & 5014 & 3293\\
104 & 191 & 351 & 662 & 1403 & 3242 & 2133\\
105 & 121 & 197 & 381 & 855 & 2084 & 1357\\
106 & 64 & 122 & 210 & 495 & 1284 & 836\\
107 & 34 & 74 & 120 & 274 & 774 & 521\\
108 & 16 & 41 & 66 & 152 & 433 & 297\\
109 & 12 & 30 & 39 & 83 & 252 & 167\\
110 & 6 & 17 & 21 & 49 & 130 & 92\\
111 & 4 & 10 & 15 & 26 & 69 & 47\\
112 & 3 & 3 & 11 & 15 & 29 & 22\\
113 & 2 & 2 & 8 & 5 & 15 & 9\\
114 & 1 & 2 & 4 & 3 & 7 & 2\\
115 & 0 & 1 & 1 & 0 & 3 & 2\\
116 & 0 & 1 & 1 & 0 & 1 & 1\\
117 & 0 & 0 & 0 & 0 & 1 & 1\\
\bottomrule
\end{tabular}
\end{table}

\hypertarget{implementation-details-and-setup}{%
\subsection{Implementation details and setup}\label{implementation-details-and-setup}}

The syntax used by the \texttt{longevity} package purposely mimics that of the \texttt{survival} package (Terry M. Therneau and Grambsch 2000). In \textbf{R}, regression models can be specified using a formula and a data frame: these are normally used in functions such as \texttt{coxph} or \texttt{survfit} in \texttt{survival}. In \texttt{longevity}, there are no regression models and many arguments are optional; they need not even be of the same length. As such, \texttt{longevity} uses instead a named list \texttt{args} to pass arguments which are typically common to all functions, to avoid having to specify them every time. For a given data, the elements in \texttt{args} will match across function as \texttt{time}, \texttt{time2}, \texttt{ltrunc}, \texttt{rtrunc}, \texttt{event}, etc., all of which characterize the data and the survival mechanisms at play. Default values are overriden by elements in \texttt{args}, with the exception of those that are passed by the user directly in the call. Relative to \pkg{survival}, additional arguments \texttt{ltrunc} and \texttt{rtrunc} for left and right truncation limits are specified, possibly matrices for the case of double interval truncation (Belzile et al. 2022), since both censoring and truncation can be present. \texttt{longevity} uses the S3 object oriented system; generics are discussed in the sequel.

To show the syntax, we build the time vectors and truncation bounds for the Dutch data. We rescale observations to years for interpretability and keep only records above age 98 for simplicity.
We split the data to handle the observed age at death first, using \texttt{interval2} to specify interval data using both time limits: these are treated as observed (uncensored) whenever \texttt{time} and \texttt{time2} coincide. When exact dates are not available, we compute the range of possible age at which individuals may have died, given their birth and death years and months.

\begin{verbatim}
library(longevity)
library(lubridate)
library(dplyr, warn.conflicts = FALSE)
data(dutch, package = "longevity")
yr_samp <- year(attr(x = dutch, which = "sampling_frame"))
dutch1 <- dutch |>
  subset(!is.na(ndays)) |>
  # Remove interval censored data for the time being
  mutate(time = ndays / 365.25, # age at death
         time2 = time,
         # min/max age to be included in sampling frame
         ltrunc = ltrunc / 365.25,
         rtrunc = rtrunc / 365.25,
         event = 1) |> # observed failure time (event=1)
  subset(time > 98) |>
  select(time, time2, ltrunc, rtrunc, event, gender, byear)
# Subset all interval-censored interval-truncated records
dutch2 <- dutch |>
  subset(is.na(ndays)) |>
  mutate(time2 = ceiling_date(
    dmy(paste("01-", dmonth, "-", dyear)), unit = "month") - 1 -
            dmy(paste("01-01-", byear)),
          time = dmy(paste("01-", dmonth, "-", dyear)) -
                 dmy(paste("31-12-", byear)),
         ltrunc = dmy(paste("01-01-1986")) - dmy(paste("31-12-", byear)),
         rtrunc = dmy(paste("31-12-2015")) - dmy(paste("01-01-", byear))
         ) |>
  select(time, time2, ltrunc, rtrunc, gender, byear) |>
  mutate(time = as.numeric(time) / 365.25, # lower censoring limit
         time2 = as.numeric(time2) / 365.25, # upper censoring limit
         ltrunc = as.numeric(ltrunc) / 365.25, # min age to be included
         rtrunc = as.numeric(rtrunc) / 365.25, # max age to be included
         event = 3) |> # interval censoring
  subset(time > 98)
# Combine databases
dutch_data <- rbind(dutch1, dutch2)
\end{verbatim}

We can proceed similarly for the Japanese data, but this time all observations are interval censored with one-year intervals. Counts are treated as weight vectors in the log likelihood.

\begin{verbatim}
data(japanese2, package = "longevity")
# Keep only non-empty cells
japanese2 <- japanese2[japanese2$count > 0, ]
# Define arguments that are recycled
japanese2$rtrunc <- 2020 -
  as.integer(substr(japanese2$bcohort, 1, 4))
# The line above etracts the earliest year of the birth cohort
args_japan <- with(japanese2,
             list(
               time = age,
               time2 = age + 1L,
               event = 3,
               type = "interval2",
               rtrunc = rtrunc,
               weights = count))
\end{verbatim}

\hypertarget{parametric-models}{%
\section{Parametric models}\label{parametric-models}}

The \texttt{longevity} package includes several parametric models commonly used in demography and extreme value theory; their hazard functions are given in Table \ref{tab:tbl-parametric-models}. Two of those models, labelled \texttt{perks} and \texttt{beard} are logistic-type hazard functions proposed in Perks (1932) that have been used by Beard (1963), and popularized in work of Kannisto and Thatcher; we use the parametrization of Richards (2012), from which we also adopt the nomenclature.

Many other \textbf{R} packages have such utilities: the \texttt{fitdistrplus} package (Delignette-Muller and Dutang 2015) has various optimization routines and allows for user-specified parametric distributions, thus handling the case of truncated distributions. Many parametric distributions also appear in the \texttt{VGAM} package (Yee and Wild 1996; Yee 2015), which allows for vector generalized linear modelling. The \texttt{survival} package can fit a range of accelerated failure time models. The \texttt{longevity} package is less general and offers support only for selected parametric distributions, but includes model comparison tools that account for nonregular asymptotic and goodness-of-fit diagnostics.

\begin{table}

\caption{\label{tab:tbl-parametric-models}Parametric models for excess lifetime}
\centering
\begin{tabular}[t]{lll}
\toprule
model & hazard function & constraints\\
\midrule
\texttt{exp} & \(\sigma^{-1}\) & \(\sigma > 0\)\\
\texttt{gomp} & \(\sigma^{-1}\exp(\beta t/\sigma)\) & \(\sigma > 0, \beta > 0\)\\
\texttt{gp} & \((\sigma + \xi t)_{+}^{-1}\) & \(\sigma > 0, \xi \in \mathbb{R}\)\\
\texttt{weibull} & \(\sigma^{-\alpha} \alpha t^{\alpha-1}\) & \(\sigma > 0, \alpha  >  0\)\\
\texttt{extgp} & \(\beta\sigma^{-1}\exp(\beta t/\sigma)[\beta+\xi\{\exp(\beta t/\sigma) -1\}]^{-1}\) & \(\sigma > 0, \beta > 0, \xi \in \mathbb{R}\)\\
\texttt{extweibull} & \(\alpha\sigma^{-\alpha}t^{\alpha-1}\{1+\xi(t/\sigma)^{\alpha}\}_{+}\) & \(\sigma > 0, \alpha > 0, \xi \in \mathbb{R}\)\\
\texttt{perks} & \(\{\alpha\exp(\nu x)\}/\{1+\alpha\exp(\nu x)\}\) & \(\nu \geq  0, \alpha >0\)\\
\texttt{beard} & \(\{\alpha\exp(\nu x)\}/\{1+\alpha\beta\exp(\nu x)\}\) & \(\nu \geq  0, \alpha >0, \beta \geq 0\)\\
\texttt{gompmake} & \(\lambda + \sigma^{-1}\exp(\beta t/\sigma)\) & \(\lambda \ge 0, \sigma > 0, \beta > 0\)\\
\texttt{perksmake} & \(\lambda + \{\alpha\exp(\nu x)\}/\{1+\alpha\exp(\nu x)\}\) & \(\lambda \geq 0, \nu \geq  0, \alpha > 0\)\\
\texttt{beardmake} & \(\lambda +  \alpha\{\exp(\nu x)\}/\{1+\alpha\beta\exp(\nu x)\}\) & \( \lambda \geq 0, \nu \ge 0, \alpha > 0, \beta \geq 0\)\\
\bottomrule
\end{tabular}
\end{table}

\begin{figure}

{\centering \includegraphics[width=0.9\textwidth]{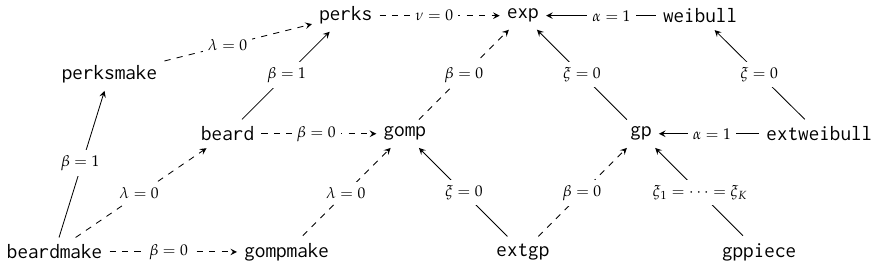}

}

\caption{Relationship between parametric models showing nested relations. Dashed arrows represent restrictions that lead to nonregular asymptotic null distribution for comparison of nested models. Comparisons between models with Makeham components and exponential are not permitted by the software because of nonidentifiability issues.}\label{fig:fig-nesting}
\end{figure}

Many of the models are nested and Figure \ref{fig:fig-nesting} shows the logical relation between the various families. The function \texttt{fit\_elife} allows users to fit all of parametric models of Table \ref{tab:tbl-parametric-models}: the \texttt{print} method returns a summary of the sampling mechanism, the number of observations, the maximum log likelihood and parameter estimates with standard errors. Depending on the data, some models may be overparametrized and parameters need not be numerically identifiable. To palliate to such issues, the optimization routine, which uses \texttt{Rsolnp}, can fit various submodels with potentially multiple starting values to ensure that the value returned are indeed the maximum likelihood estimates. Calls to \texttt{anova} where this isn't the case return errors.

The \texttt{fit\_elife} function handles arbitrary censoring patterns over single intervals, along with single interval truncation an interval censoring. To accommodate the sampling scheme of the International Database on Longevity, an option also allows for double interval truncation (Belzile et al. 2022), whereby observations are included only if the person dies between time intervals, potentially overlapping, which defines the observation window over which death individuals are recorded. To avoid large data storage with identical observations which are often provided in the form of tabulated data, an additional \texttt{weights} argument can be provided.

\begin{verbatim}
thresh <- 108
model0 <- fit_elife(arguments = args_japan,
                    thresh = thresh,
                    family = "exp")

(model1 <- fit_elife(arguments = args_japan,
                      thresh = thresh,
                      family = "gomp"))
\end{verbatim}

\begin{verbatim}
#> Model: Gompertz distribution.
#> Sampling: interval censored, right truncated
#> Log-likelihood: -3599.037
#>
#> Threshold: 108
#> Number of exceedances: 2489
#>
#> Estimates
#>  scale   shape
#> 1.6855  0.0991
#>
#> Standard Errors
#>  scale   shape
#> 0.0523  0.0273
#>
#> Optimization Information
#>   Convergence: TRUE
\end{verbatim}

\hypertarget{model-comparisons}{%
\subsection{Model comparisons}\label{model-comparisons}}

Goodness of fit of nested models can be compared using likelihood ratio tests via the \texttt{anova} method. Most of the interrelations between models yield non-regular model comparisons since, to recover the simpler model, one must often fix one or more parameter to values that lie on the boundary of the parameter space. For example, if we compare a Gompertz model with the exponential, the limiting null distribution is a mixture of a point mass at zero and a \(\chi^2_1\) variable, both with probability half (Chernoff 1954). Many authors (e.g., Camarda 2022) fail to recognize this fact. The case becomes more complicated with more than one boundary constraint: for example, the deviance statistic comparing the Beard--Makeham and the Gompertz model, which constrains two parameters on the boundary of the parameter space, has a null distribution which is a mixture of \(\frac{1}{4}\chi^2_2 + \frac{1}{2}\chi^2_1 + \frac{1}{4}\chi^2_0\) (Self and Liang 1987).

Nonidentifiability also impacts testing: for example, if the rate parameter of the Perks--Makeham model (\texttt{perksmake}) \(\nu \to 0\), the limiting hazard is \(\lambda + \exp(\alpha)/\{1+\exp(\alpha)\}\) and constant, corresponding to the exponential model. However, neither \(\alpha\) nor \(\lambda\) is identifiable. The usual asymptotics for the likelihood ratio test break down as the information matrix is singular (Rotnitzky et al. 2000). All three families that include a Makeham component cannot be directly compared to the exponential in \texttt{longevity} and the call to \texttt{anova} returns an error message.

Below, we test whether the exponential model is an adequate simplification of the Gompertz model for exceedances above 108 years. The drop in log likelihood is quite large, indicating the exponential model is not an adequate simplification of the Gompertz fit.

\begin{verbatim}
# Model comparison
anova(model1, model0)
\end{verbatim}

\begin{tabular}{lrrrrr}
\toprule
  & npar & Deviance & Df & Chisq & Pr(>Chisq)\\
\midrule
gomp & 2 & 7198.07 &  &  & \\
exp & 1 & 7213.25 & 1 & 15.17 & 0\\
\bottomrule
\end{tabular}

\hypertarget{sampling}{%
\subsection{Sampling}\label{sampling}}

Given the poor finite sample properties of the aforementioned tests, it may be preferable to rely on a parametric bootstrap rather than on the asymptotic distribution of the test statistic (Belzile et al. 2022).
Simulation-based inference requires capabilities for drawing new datasets whose features match that of the original one. For example, the \href{supercentenarians.org}{International Database on Longevity} (IDL) (Jdanov, Shkolnikov, and Gellers-Barkmann 2021) features data that are interval truncated above 110 years, but doubly interval truncated since the sampling period for semisupercentenarians (who died age 105 to 110) and supercentenarians (who died above 110) are not always the same (Belzile et al. 2022). The Istat 2018 database analyzed by Barbi et al. (2018) on Italian includes left truncated right censored records.

To mimic the postulated data generating mechanism while accounting for the sampling scheme, we could use the observed birth dates, or simulate new birth dates (possibly through a kernel estimator of the empirical distribution of birth dates) while keeping the sampling frame with the first and last date of data collection to define the truncation interval. In other settings, one could obtain the nonparametric maximum likelihood estimator of the distribution of the upper truncation bound (Shen 2010) using an inverse probability weighted estimator, which for fixed data collection windows is equivalent to setting the birth date.

The \texttt{samp\_elife} function includes multiple \texttt{type2} arguments to handle these. For interval truncated data (\texttt{type2="ltrt"}), it uses the inversion method (Section 2 of Devroye (1986)): for \(F\) an absolutely continuous distribution function and \(F^{-1}\) the corresponding quantile function, a random variable distributed according to \(F\) with truncated on \([a,b]\) is generated as \(X \sim F^{-1}[F(a) + U\{F(b)-F(a)\}]\)
where \(U \sim \mathsf{U}(0,1)\) is standard uniform.

The function \texttt{samp\_elife} includes an argument \texttt{upper} which serves for both right truncation, and right censoring. For the latter, any record simulated that exceeds \texttt{upper} is capped at that upper bound and declared partially observed. This is useful for simulating administrative censoring, whereby the birth date and the upper bound of the collection window fully determine whether an observation is right censored or not. An illustrative example is provided in the next section.

\hypertarget{simulation-based-inference}{%
\subsubsection{Simulation-based inference}\label{simulation-based-inference}}

The \texttt{anova} call uses the asymptotic null distribution for comparison of nested parametric distributions \(\mathcal{F}_0 \subseteq \mathcal{F}_1\). We could use the bootstrap to see how good this approximation to the null distribution is. To mimic as closely as possible the data generating mechanism, which is custom in most scenarios, we condition on the sampling frame and the number of individuals in each birth cohort. The number dying at each age is random, but the right truncation limits will be the same for anyone in that cohort. We simulate excess lifetimes, then interval censor observations by keeping only the corresponding age bracket. Under the null hypothesis, the data are drawn from \(\widehat{F}_0 \in \mathcal{F}_0\) and we generate observations from this right truncated distribution using the \texttt{samp\_elife} utility, which also supports double interval truncation and left truncation right censoring. This must be done within a for loop since we have count attached to each upper bound, but the function is vectorized should we use a single vector containing all of the right truncation limits.

The bootstrap \(p\)-value for comparing models \(M_0 \subset M_1\) would be obtained by repeating the following steps \(B\) times and calculating the rank of the observed test statistic among alternatives:

\begin{enumerate}
\def\labelenumi{\arabic{enumi}.}
\tightlist
\item
  Simulate new birth dates \(d_i\) \((i=1, \ldots, n)\) (e.g., drawing from a smoothed the empirical distribution of birth dates); the latest possible birth date is one which ensures the person reached at least the threshold by the end of the period.
\item
  Subtract the endpoints of the sampling period, say \(c_1\) and \(c_2\) to get the minimum (maximum) age at death, \(c_1 - d_i\) (respectively \(c_2 - d_i\)) days, which define the truncation bounds.
\item
  Use the function \texttt{samp\_elife} to simulate new observations from a parametric interval truncated distribution from the null model \(M_0\)
\item
  Use the optimization procedure in \texttt{fit\_elife} to fit the model with both \(M_1\) and \(M_2\) and calculate the deviance or likelihood ratio statistic.
\end{enumerate}

The algorithm is implemented below for comparing the Gompertz and the exponential model.

\begin{verbatim}
set.seed(2022)
# Count # of unique right truncation limit
db_rtrunc <- aggregate(count ~ rtrunc,
                FUN = "sum",
                data = japanese2,
                subset = age  >= thresh)
# Number of bootstrap replications
B <- 9999L
boot_anova <- numeric(length = B + 1L)
boot_gof <- numeric(length = B + 1L)
for(b in seq_len(B)){
  # Generate bootstrap sample
  boot_samp <-
    do.call(rbind, #merge data frames
     apply(db_rtrunc, 1, function(x){
    # for each rtrunc and count
    count <- table( #tabulate count
      floor( #round down
      # sample right truncated exponential
      samp_elife(n = x["count"],
             scale = model0$par,
             family = "exp", #null model
             upper = x["rtrunc"] - thresh,
             type2 = "ltrt")))
    # return data frame
    data.frame(count = as.integer(count),
               rtrunc = as.numeric(x["rtrunc"]) - thresh,
               eage = as.integer(names(count)))
  }))
  # Fit null model to bootstrap sample
  boot_mod0 <-
    with(boot_samp,
         fit_elife(time = eage,
            time2 = eage + 1L,
            rtrunc = rtrunc,
            type = "interval",
            event = 3,
            family = "exp",
            weights = count))
  # Fit alternative model to bootstrap sample
  boot_mod1 <-
    with(boot_samp,
         fit_elife(time = eage,
            time2 = eage + 1L,
            rtrunc = rtrunc,
            type = "interval",
            event = 3,
            family = "gomp",
            weights = count))
  boot_anova[b] <- deviance(boot_mod0) -
    deviance(boot_mod1)
}
# Add original statistic
boot_anova[B] <- deviance(model1) - deviance(model0)
# Bootstrap p-value
boot_pval <- rank(boot_anova)[B] / B
\end{verbatim}

The asymptotic approximation is of similar magnitude as the bootstrap \(p\)-value, which is 2\(\times 10^{-4}\). Both suggest that the more complex Gompertz model provides a significantly better fit.

\hypertarget{extreme-value-analysis}{%
\subsection{Extreme value analysis}\label{extreme-value-analysis}}

The \texttt{longevity} package was initially built for dealing with records of supercentenarians, i.e., people who died above age 110. Given the rarity of such individuals, parametric models originating from extreme value analysis are often justified. Univariate extremes is well implemented in software; Belzile, Dutang, et al. (2023) provides a recent review of implementations. While there are many standard R packages for the analysis of univariate extremes using likelihood-based inference, such as \texttt{evd} (Stephenson 2002), \texttt{mev} and \texttt{extRemes} (Gilleland and Katz 2016), only the \texttt{evgam} package includes functionalities to fit threshold exceedance models with censoring, as showcased in Youngman (2022) with rounded rainfall measurements.

Study of population dynamics and mortality generally requires knowledge of the total population from which observations are drawn to derive rates. By contrast, the peaks over threshold method, by which one models the \(k\) largest observations of a sample, is a conditional analysis (e.g., given survival until a certain age), and is therefore free of denominator specification since we only model exceedances above a high threshold \(u\). Extreme value theory suggests that, in many instances, the limiting conditional distribution of exceedances of a random variable \(Y\) with distribution function \(F\) is generalized Pareto, meaning
\begin{align}
\lim_{u \to x^*}\Pr(Y-u  > y \mid Y  > u)= \begin{cases}
\left(1+\xi y/\sigma\right)_{+}^{-1/\xi}, & \xi \neq 0;\\
\exp(-y/\sigma), & \xi = 0;
\end{cases}
\label{eq:gpd}
\end{align}
with \(x_{+} = \max\{x, 0\}\) and \(x^*=\sup\{x: F(x) < 1\}\). This justifies the use of Equation \eqref{eq:gpd} for the survival function of threshold exceedances when dealing with rare events. The model has two parameters: a scale \(\sigma\) and a shape \(\xi\) which determines the behaviour of the upper tail. Negative shape parameters correspond to bounded upper tails and a finite right endpoint for the support.

\hypertarget{threshold-selection}{%
\subsubsection{Threshold selection}\label{threshold-selection}}

For modelling purposes, we need to pick a threshold \(u\) that is smaller than the upper endpoint \(x^*\) in order to have sufficient number of observations to estimate parameters. The threshold selection problem is a classical instance of bias-variance trade-off: the parameter estimators are possibly biased if the threshold is too low because the generalized Pareto approximation is not good enough, whereas choosing a larger threshold to ensure we are closer to the asymptotic regime leads to reduced sample size and increased parameter uncertainty.

To aid selection, we resort notably to threshold stability plots. These are common visual diagnostics consisting of a plot of estimates of the shape parameter \(\widehat{\xi}\) (with confidence or credible intervals) based on sample exceedances over a range of thresholds \(u_1, \ldots, u_K\). If the data were drawn from a generalized Pareto distribution, the conditional distribution above higher threshold \(v >u\) is also generalized Pareto with the same shape: this threshold stability property is the basis for extrapolation beyond the range of observed records. If the change in estimation of \(\xi\) is nearly constant, this provides reassurance that the approximation can be used for extrapolation. The only difference with survival data, relative to the classical setting, is that the likelihood must account for censoring and truncation. Note that, when we use threshold exceedances with a nonzero threshold (argument \texttt{thresh}), it however ain't possible to unambiguously determine whether left censored observations are still exceedances: such cases yield errors in the functions.

Theory on penultimate extremes suggests that, for finite levels and general distribution function \(F\) for which \eqref{eq:gpd} holds, the shape parameter varies as a function of the threshold \(u\), behaving like the derivative of the reciprocal hazard \(r(x) = \{1-F(x)\}/f(x)\). We can thus model the shape as a piecewise constant by fitting a piecewise generalized Pareto model due to Northrop and Coleman (2014) and adapted in Belzile et al. (2022) for survival data. The latter can be viewed as a mixture of generalized Pareto over \(K\) disjoint intervals (with continuity constraints to ensure a smooth hazard) which reduces to the generalized Pareto if we force the \(K+1\) shape parameters to be equal. We can use a likelihood ratio tests to compare the model, or a score test if the latter is too computationally intensive, and plot the \(p\)-values for each of the \(K\) threshold, corresponding to the null hypotheses \(\mathrm{H}_k: \xi_k = \cdots = \xi_{K}\) (\(k=1, \ldots, K-1\)). As the model quickly becomes overparametrized, optimization is difficult and the score test may be a safer option as it only requires estimation of the null model of a single generalized Pareto over the whole range.

\begin{figure}

{\centering \includegraphics[width=0.8\linewidth]{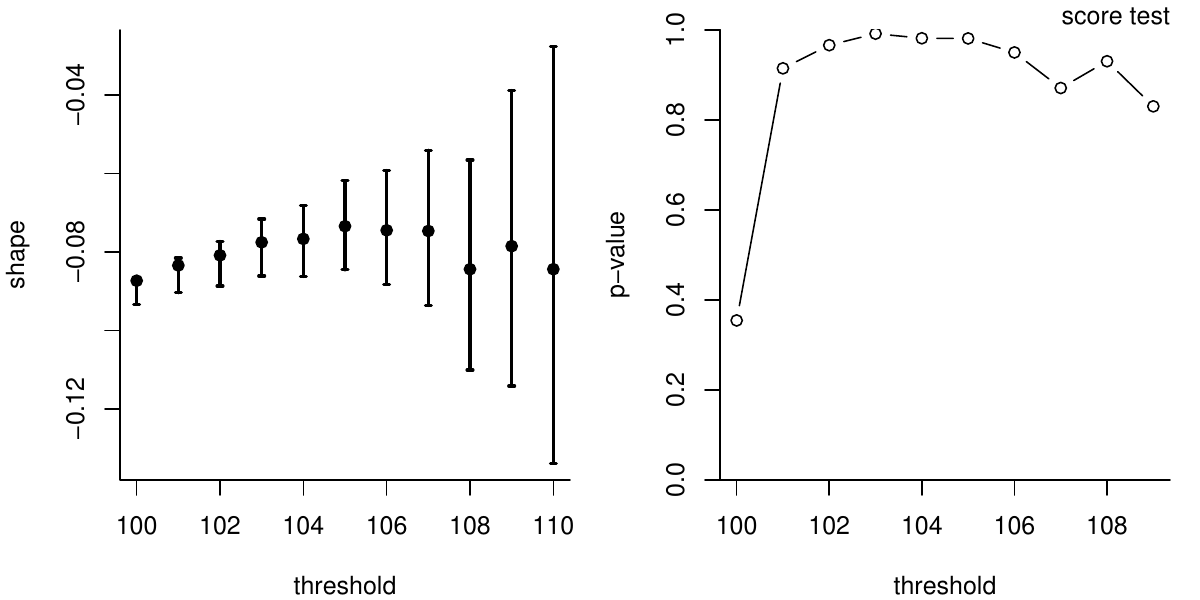}

}

\caption{Threshold diagnostic tools: parameter stability plots for the generalized Pareto model (left) and Northrop--Coleman \(p\)-value path (right) for the Japanese centenarian dataset.}\label{fig:fig-parameterstab}
\end{figure}

To illustrate these diagnostics tools, Figure \ref{fig:fig-parameterstab} shows a threshold stability plot, which features a small increase in the shape parameters as the threshold increase, corresponding to a stabilization or even a slight decrease of the hazard at higher ages. We can envision a threshold of 108 years as being reasonable: the Northrop--Coleman diagnostic plots suggests lower threshold are compatible with a constant shape above 100. Additional goodness-of-fit diagnostics are necessary to determine if the generalized Pareto model fits well.

\begin{verbatim}
par(mfrow = c(1, 2),
    mar = c(4, 4, 1, 1))
# Threshold sequence
u <- 100:110
# Threshold stability plot
do.call(
  what = tstab,
  args = c(args_japan,
           list(family = "gp",
                method = "profile",
                which.plot = "shape",
                thresh = u)))
# Northrop-Coleman diagnostic based on score tests
nu <- length(u) - 1L
nc_score <-
  do.call(what = nc_test,
        args = c(args_japan,
                 list(thresh = u)))
score_plot <- plot(nc_score)
graphics.off()
\end{verbatim}

Each plot in the package can be produced using base R or using \texttt{ggplot2} (Wickham 2016), which implements the grammar of graphics. To keep the list of package dependencies lean and adhere to the \href{https://cran.r-project.org/web/packages/pacs/vignettes/tinyverse.html}{\texttt{tinyverse} principle}, the latter can be obtained by using the argument \texttt{plot.type} with the generic S3 method \texttt{plot}, rather than via \texttt{autoplot}, provided the package is already installed.

\hypertarget{graphical-goodness-of-fit-diagnostics}{%
\subsection{Graphical goodness-of-fit diagnostics}\label{graphical-goodness-of-fit-diagnostics}}

Determining whether a parametric model fits well to survival data is no easy task due to the difficulty in specifying the null distribution of many goodness-of-fit statistics, such as the Cramér--von Mises statistic, which differ for survival data. As such, the \texttt{longevity} package relies mostly on visual diagnostic tools. Waller and Turnbull (1992) discusses how classical visual graphics can be adapted in the presence of censoring: only observed failure times, shown on the \(y\)-axis are displayed against their empirical plotting position on the \(x\)-axis. Contrary to the independent and identically distributed case, the uniform plotting positions \(F_n(y_i)\) are based on the nonparametric maximum likelihood estimator discussed previously.

The situation is more complicated with truncated data (Belzile et al. 2022), since the data are not identically distributed: indeed, the distribution function of observation \(Y_i\) truncated on the interval \([a_i, b_i]\) is \(F_i(y_i) = \{F(y_i)-F(a_i)\}/\{F(b_i) - F(a_i)\}\), so the data arise from different distributions even if these share common parameters. One way out of this conundrum is using the probability integral transform and the quantile transform to map observations to the uniform scale and back onto the data scale. Taking \(\widetilde{F}(y_i) = F_n(y_i)=\mathrm{rank}(y_i)/(n+1)\) to denote the empirical distribution function estimator, a probability-probability plot would show \(x_i = \widetilde{F}_{i}(y_i)\) against \(y_i = F_i(y_i)\), leading to approximately uniform samples if the parametric distribution \(F\) is suitable. Another option is to standardize the observation, taking the collection \(\widetilde{y}_i=F^{-1}\{F_i(y_i)\}\) of rescaled exceedances and comparing them to the usual plotting positions \(x_{(i)} =\{i/(n+1)\}\). The drawback of the latter approach is that the quantities displayed on the \(y\)-axis are not raw observations and the ranking of the empirical quantiles may change, a somewhat counterintuitive feature. However, this means that the sample \(\{F_i(y_i)\}\) should be uniform under the null hypothesis, and this allows one to use methods from Säilynoja, Bürkner, and Vehtari (2022) to obtain pointwise and simultaneous confidence intervals.

\texttt{longevity} offers users the choice between quantile-quantile (Q-Q, \texttt{"qq"}) and Tukey's mean detrended Q-Q plots (\texttt{"tmd"}) and exponential Q-Q plots (\texttt{"exp"}). Other options on uniform scale are probability-probability (P-P, \texttt{"pp"}) plots and empirically rescaled plots (ERP, \texttt{"erp"}) (Waller and Turnbull 1992), designed to ease interpretation with censored observations by rescaling axes. We illustrate the graphical tools with the \texttt{dutch} data above age 105.

\begin{verbatim}
fit_dutch <- fit_elife(
   arguments = dutch_data,
   event = 3,
   type = "interval2",
   family = "gp",
   thresh = 105,
   export = TRUE)
par(mfrow = c(1, 2))
plot(fit_dutch,
     which.plot = c("pp","qq"))
\end{verbatim}

\begin{figure}

{\centering \includegraphics[width=1\linewidth]{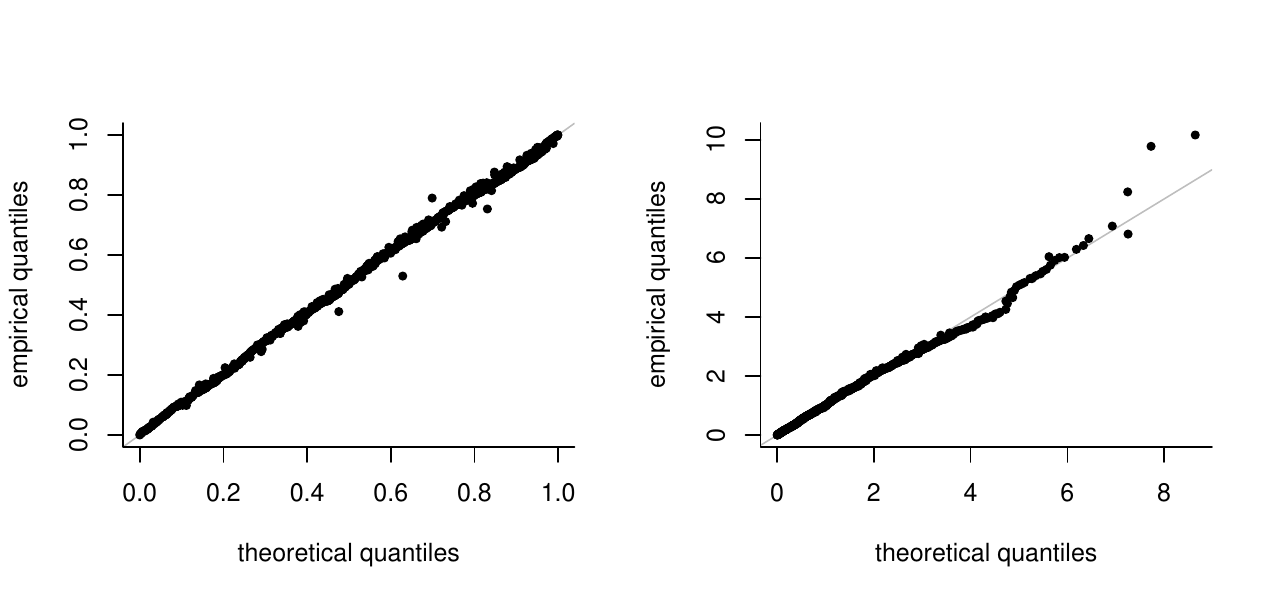}

}

\caption{Probability-probability and quantile-quantile plots for generalized Pareto model fitted above age 105 years to Dutch data.}\label{fig:fig-qqplots}
\end{figure}

Censored observations are used to compute the plotting positions, but are not displayed on visual diagnostics. As such, we cannot use graphical goodness of fit diagnostics for the Japanese interval censored data. An alternative, given that data are tabulated in a contingency table, is to use a chi-squared test for independence, conditioning on the number of individuals per birth cohort. The expected number in each cell (birth cohort and age band) can be obtained by computing the conditional probability of falling in that age band. The asymptotic null distribution should be \(\chi^2\) with \((k-1)(p-1)\) degrees of freedom, where \(k\) is the number of age bands and \(p\) the number of birth cohorts. In finite samples, the expected count for large excess lifetimes are very low so one can expect the \(\chi^2\) approximation to be poor. To mitigate this, we can pool observations and resort to simulation to approximate the null distribution of the test statistic. The bootstrap \(p\)-value for the exponential model above 108 years, pooling observations with excess lifetime of 5 years and above, is 0.859, indicating no evidence that the model is inadequate, but the test here may have low power.

\hypertarget{stratification}{%
\subsection{Stratification}\label{stratification}}

Demographers may suspect differences between individuals of different sex, from different countries or geographic area, or by birth cohort. All of these are instances of categorical covariates. One possibility is to incorporate these covariates with suitable link function through parameters, but we consider instead stratification. We can split the data by levels of \texttt{covariate} (with factors) into sub-data and compare the goodness-of-fit of the \(K\) models relative to that which pools all observations. The \texttt{test\_elife} function performs likelihood ratio tests for the comparisons. We illustrate this with a generalized Pareto model for the excess lifetime.

\begin{verbatim}
test_elife(arguments = args_japan,
           thresh = 110,
           family = "gp",
           covariate = japanese2$gender)
\end{verbatim}

The null hypothesis is \(\mathrm{H}_0: \sigma_{\texttt{f}} = \sigma_{\texttt{m}}, \xi_{\texttt{f}}=\xi_{\texttt{m}}\) against the alternative that at least one equality doesn't hold and so the hazard and endpoints are different.
In the present example, there is no evidence against any difference in lifetime distribution between male and female; this is unsurprising given the large imbalance between count of each covariate level, with much fewer male than female.

\hypertarget{extrapolation}{%
\subsection{Extrapolation}\label{extrapolation}}

If the maximum likelihood estimator of the shape \(\xi\) for the generalized Pareto model is negative, then the distribution has a finite upper endpoint; otherwise, the latter is infinite. With \(\xi < 0\), we can look at the profile log likelihood for the endpoint \(\eta = -\sigma/\xi\), using the function \texttt{prof\_gp\_endpt}, to draw the curve and obtain confidence intervals. The argument \texttt{psi} is used to give a grid of values over which to compute the profile log likelihood. The bounds of the confidence intervals are obtained by fitting a cubic smoothing spline for \(y=\eta\) as a function of the shifted profile curve \(x = 2\{\ell_p(\eta)-\ell_p(\widehat{\eta})\}\) on both side of the maximum likelihood estimator and predicting the value of \(y\) when \(x = -\chi^2_1(0.95)\). This technique works well unless the profile is nearly flat or the bounds lie beyond the range of values of \texttt{psi} provided; the user may wish to change them if they are too far. If \(\widehat{\xi} \approx 0\), then the upper bound of the confidence interval may be infinite and the profile log likelihood may never reach the cutoff value of the asymptotic \(\chi^2_1\) distribution.

The profile log likelihood curve for the endpoint, shifted so that it's value is zero at the maximum likelihood estimator, highlights the marked asymmetry of the distribution of \(\eta\), with the horizontal dashed lines showing the limits for the 95\% profile likelihood confidence intervals. These suggest that the endpoint, or a potential finite lifespan, could lie very much beyond observed records. The routine used to calculate the upper bound computes the cutoff value by fitting a smoothing spline with the role of the \(y\) and \(x\) axes reverses and by predicting the value of \(\eta\) at \(y=0\). In this example, the upper confidence limit is extrapolated from the model: more accurate measure can be obtained by specifying a longer and finer sequence of values of \texttt{psi} such that the profile log likelihood drops below the \(\chi^2_1\) quantile cutoff.

\begin{verbatim}
# Create grid of threshold values
thresholds <- 105:110
# Grid of values at which to evaluate profile
psi <- seq(120, 200, length.out = 101)
# Calculate the profile for the endpoint
# of the generalized Pareto at each threshold
profiles <- lapply(thresholds, function(thresh){
  do.call(prof_gp_endpt,
          args = c(args_japan, list(thresh = thresh, psi = psi)))
})
# Compute corresponding confidence intervals
conf <- data.frame(threshold = thresholds,
                   t(sapply(profiles, confint)))
# Plot point estimates and confidence intervals
g1 <- ggplot(data = conf,
       mapping = aes(x = threshold,
                     y = Estimate,
                     ymin = Lower.CI,
                     ymax = Upper.CI)) +
  geom_pointrange() +
  labs(y = "lifespan (in years)") +
  theme_classic()
# Plot the profile curve with cutoffs for conf. int. for 110
g2 <- plot(profiles[[6]], plot.type = "ggplot", plot = FALSE)
patchwork::wrap_plots(g1, g2)
\end{verbatim}

\begin{figure}

{\centering \includegraphics[width=0.9\linewidth]{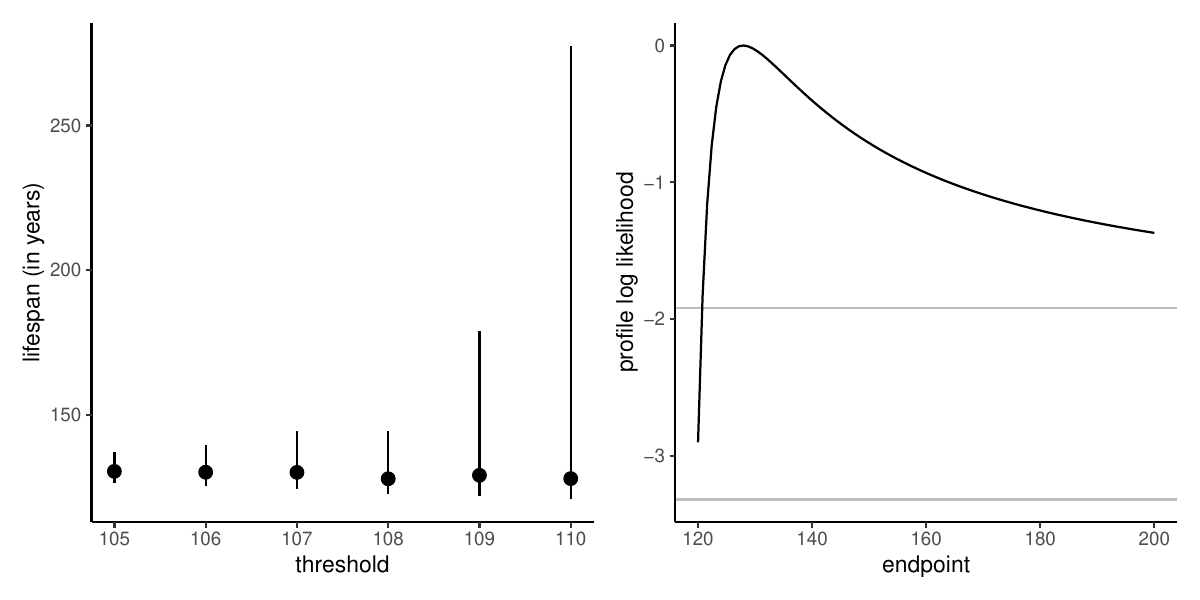}

}

\caption{Maximum likelihood estimates with 95\% confidence intervals as a function of threshold (left) and profile likelihood for exceedances above 110 years (right) for Japanese centenarian data.}\label{fig:fig-endpoint-confint}
\end{figure}

Depending on the model, the conclusions about the risk of mortality change drastically: the Gompertz model implies a ever increasing hazard, but no finite endpoint for the distribution of exceedances. The exponential model implies a constant hazard and no endpoint. By contrast, the generalized Pareto can accommodate both finite and infinite endpoints. The marked asymmetry of the distribution of lifespan defined by the generalized Pareto shows that inference obtained using symmetric confidence interval (i.e., Wald-based) is likely very misleading: the drop in fit from having a zero or positive shape parameter \(\xi\) seemingly smaller than the cutoffs for a 95\% confidence interval, suggesting that while the best point estimate is around 128 years, the upper bound is so large (and extrapolated) suggest that everything is possible. The model however also suggest a very high probability of dying in any given year, regardless of whether the hazard is constant, decreasing or increasing.

\hypertarget{hazard}{%
\subsection{Hazard}\label{hazard}}

The parameters of the models are seldom of interest in themselves: rather, we may be interested in a summary such as the hazard function. At current, \texttt{longevity} does not allow general linear modelling of model parameters or time-varying covariates, but other software implementation can tackle this task. For example, \texttt{casebase} (Bhatnagar et al. 2022) fits flexible hazard models using logistic or multinomial regression with potential inclusion of penalties for the parameters associated to covariates and splines effects. Another alternative is \texttt{flexsurv} (Jackson 2016), which has 10 parametric models and allows for user-specified models.
The \texttt{bshazard} package (Rebora, Salim, and Reilly 2014) provides nonparametric smoothing via \(B\)-splines, whereas \texttt{muhaz} handles kernel-based hazard for right-censored data; both could be used for validation of the parametric models in the case of right-censoring. \texttt{rstpm2} (Liu, Pawitan, and Clements 2018) handles generalized modelling for censoring with the Royston--Parmar model built from natural cubic splines (Royston and Parmar 2002).

Contrasting with all of the aforementioned approaches, we focus on parametric models: This is partly because there are few observations for the user-case we consider and covariates, except perhaps for gender and birth year, are not available.

\begin{figure}
\includegraphics[width=0.9\linewidth]{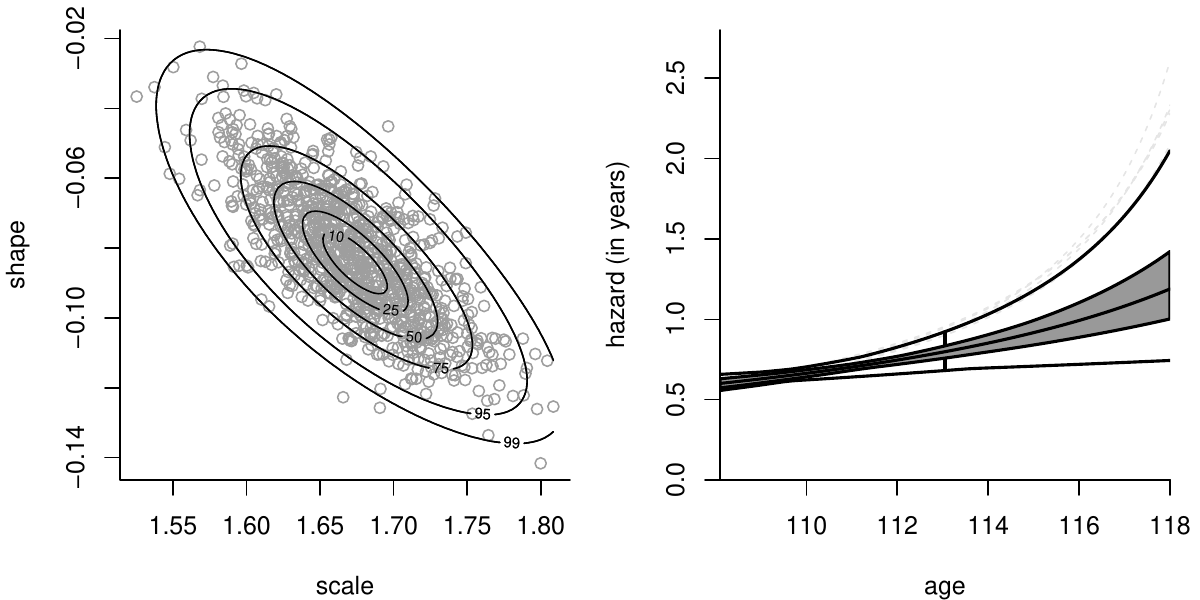} \caption{Left: scatterplot of 1000 independent posterior samples from generalized Pareto model with maximum data information prior; the contour curves give the percentiles of credible intervals. Right: functional boxplots for the corresponding hazard curves.}\label{fig:fig-hazard}
\end{figure}

Outside of the exponential hazard, which is constant and can be easily profiled, the hazard changes over time. \texttt{longevity} includes utilities for computing the hazard function from a fitted model object and computing point-wise confidence intervals using symmetric Wald intervals or the profile likelihood.
Specifically, the \texttt{hazard\_elife} function calculates the hazard \(h(t; \boldsymbol{\theta})\) point-wise at times \(t=\)\texttt{x}; Wald-based confidence intervals are obtained using the delta-method, whereas profile likelihood intervals are obtained by reparametrizing the model in terms of \(h(t)\) for each time \(t\). More naturally perhaps, we can consider a Bayesian analysis of the Japanese excess lifetime above 108 years. Using the likelihood and encoded log posterior provided in \texttt{logpost\_elife}, we obtained independent samples from the posterior of the generalized Pareto parameters \((\sigma, \xi)\) with maximal data information prior using the \texttt{rust} package. Each parameter combination was then fed into \texttt{helife} and the hazard evaluated over a range of values. Figure \ref{fig:fig-hazard} shows the posterior samples and functional boxplots (Sun and Genton 2011) of the hazard curves, obtained using the \texttt{fda}. The increase in the risk of dying at old age comes with very large uncertainty, as evidenced by the width of the boxes.

\hypertarget{nonparametric-maximum-likelihood-estimation}{%
\section{Nonparametric maximum likelihood estimation}\label{nonparametric-maximum-likelihood-estimation}}

Nonparametric methods are popular tools for the analysis of survival data, owing to their limited set of assumptions. Without explanatories, a closed-form estimator of the nonparametric maximum likelihood estimator of the survival function can be derived in particular instances, including the product limit estimator (Kaplan and Meier 1958) for the case of random or noninformative right censoring and an extension allowing for left truncation (Tsai, Jewell, and Wang 1987). In general, the nonparametric maximum likelihood estimator of the survival function needs to be computed using an expectation-maximization (EM) algorithm (Turnbull 1976). Nonparametric estimators only assign probability mass on observed failure times and intervals and so cannot be used for extrapolation beyond the range of the data.

The CRAN Task View on Survival Analysis lists various implementations of nonparametric maximum likelihood estimators of the survival or hazard functions: \texttt{survival} implements the Kaplan--Meier and Nelson--Aalen estimators. Many packages focus on the case of interval censoring (Groeneboom and Wellner 1992, 3.2); Anderson-Bergman (2017a) reviews the performance of the implementations in notably \texttt{icenReg} (Anderson-Bergman 2017b) and \texttt{Icens}. For interval truncated data, dedicated algorithms that combine gradient-based steps (Efron and Petrosian 1999) or use inverse probability weighting (Shen 2010) exist and can be more efficient than the EM algorithm of Turnbull (1976) : many of these are implemented in the \texttt{DTDA} package. The \texttt{interval} (Fay and Shaw 2010) implements Turnbull's EM algorithm for interval censored data. With a small number of observations, it is also relatively straightforward to maximize the log likelihood for the concave program subject to linear constraint using augmented Lagrangian methods and sequential quadratic programming (Ye 1987), as implemented in, for example, \texttt{Rsolnp}.

The nonparametric maximum likelihood estimator is unique only up to equivalence classes. The data for individual \(i\) consists of the tuple \(\{L_i, R_i, V_i, U_i\}\), where the censoring interval is \([L_i, R_i]\) and the truncation interval is \([V_i, U_i]\), with \(0 \leq V_i \leq L_i \leq R_i \leq U_i \leq \infty\). Turnbull (1976) shows how one can build disjoint intervals \(C = \bigsqcup_{j=1}^m [a_j, b_j]\) where \(a_j \in \mathcal{L} = \{L_1, \ldots, L_n\}\) and \(b_j \in \mathcal{R} = \{R_1, \ldots, R_n\}\) satisfy \(a_1 \leq b_1 < \cdots < a_m \leq b_m\) and the intervals \([a_j, b_j]\) contain no other members of \(L\) or \(R\) except at the endpoints. This last condition notably ensures that the intervals created include all observed failure times as singleton sets in the absence of truncation. Other authors (Lindsey and Ryan 1998) have taken interval censored data as semi-open intervals \((L_i, R_i]\), a convention we adopt here for numerical reasons. For interval censored and truncated data, Frydman (1994) shows that this construction must be amended by taking instead \(a_j \in \mathcal{L} \cup \{U_1, \ldots, U_n\}\) and \(b_j \in \mathcal{R} \cup \{V_1, \ldots, V_n\}\).

We assign probability \(p_j = F(b_j^{+}) - F(a_j^{-}) \geq 0\) to each of the resulting \(m\) intervals under the constraint \(\sum_{j=1}^m p_j = 1\) and \(p_j \geq 0\) \((j=1, \ldots, m)\). The nonparametric maximum likelihood estimator of the distribution function \(F\) is then
\begin{align*}
\widehat{F}(t) = \begin{cases} 0, & t < a_1;\\
\widehat{p}_1 + \cdots + \widehat{p}_j, & b_j < t < a_{j+1} \quad (1 \leq j \leq m-1);\\
1, & t  > b_m;
\end{cases}
\end{align*}
and is undefined for \(t \in [a_j, b_j]\) \((j=1, \ldots, m)\): users can choose to interpolate, take the left or right limit.

\begin{figure}

{\centering \includegraphics[width=0.8\linewidth]{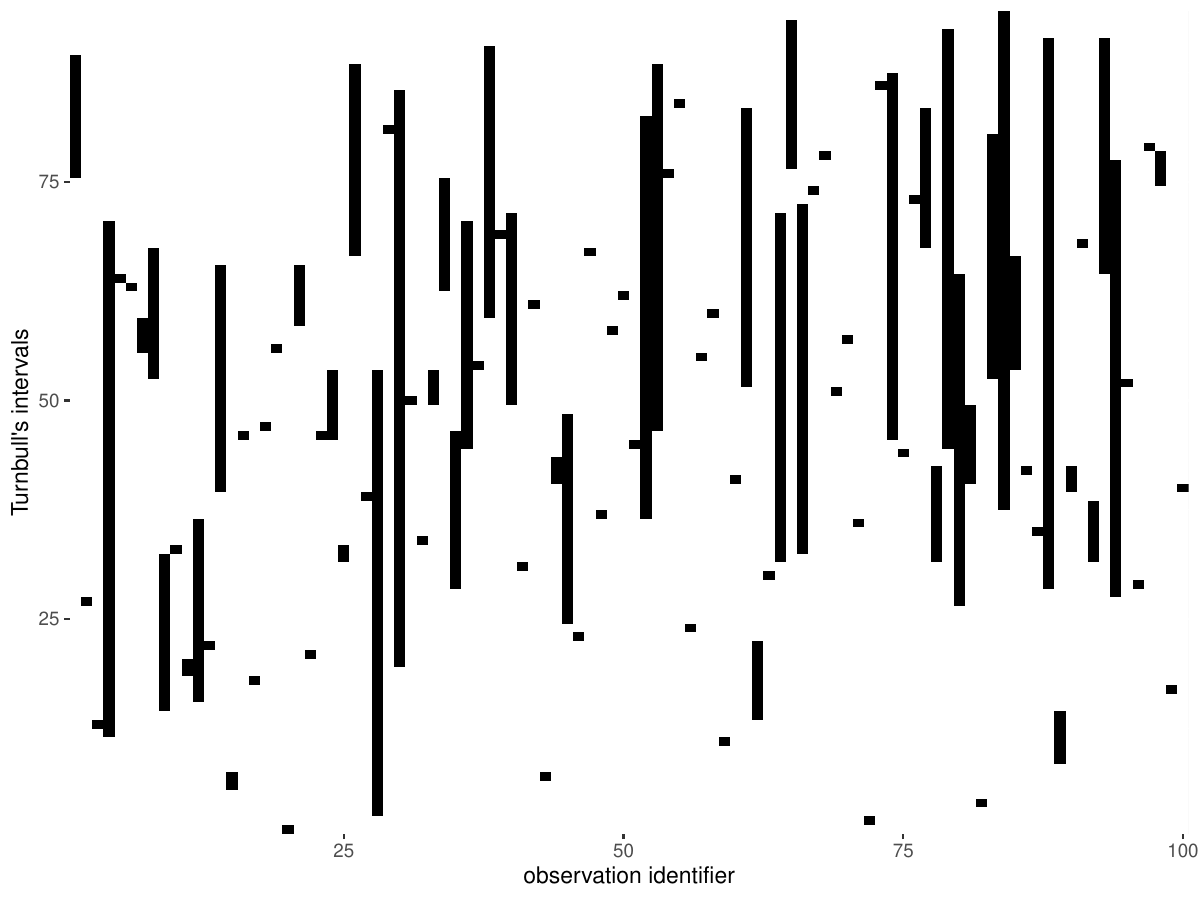}

}

\caption{Illustration of the relationship between censoring intervals and disjoint Turnbull's sets.}\label{fig:fig-turnbull}
\end{figure}

The procedure of Turnbull can be encoded using \(m \times n\) matrices. For censoring, we build \(\mathbf{A}\) whose \((i,j)\)th entry \(\alpha_{ij}=1\) if \([a_j, b_j] \subseteq A_i\) and zero otherwise. Since the intervals forming \(C\) are disjoint and in increasing increasing order, a more storage efficient manner of keeping track of the intervals is to find the smallest integer \(j\) such that \(L_i \leq a_j\) and the largest \(R_i \geq b_j\) \((j=1, \ldots, m)\) for each observation. The same idea applies for the truncation sets \(B_i = (V_i, U_i)\) and matrix \(\mathbf{B}\) with \((i,j)\) element \(\beta_{ij}\).

The log likelihood function is
\begin{align*}
\ell(\boldsymbol{p}) = \sum_{i=1}^n w_i \log \left( \sum_{j=1}^m \alpha_{ij}p_j\right) - w_i\log \left( \sum_{j=1}^m \beta_{ij}p_j\right)
\end{align*}

The numerical implementation of the EM is in principle forward: first identify the equivalence classes \(C\), next calculate the entries of \(A_i\) and \(B_i\) (or the vectors of ranges) and finally run the EM algorithm. In the second step, we need to account for potential ties in the presence of (interval) censoring and treat the intervals as open on the left for censored data. For concreteness, consider the toy example \(\boldsymbol{T} =(1,1,2)\) and \(\boldsymbol{\delta} = (0,1,1)\), where \(\delta_i = 1\) if the observation is a failure time and \(\delta_i=0\) in case of right censoring.
The left and right censoring bounds are \(\mathcal{L} = \{1, 2\}\) and \(\mathcal{R} = \{1, 2, \infty\}\) with \(A_1 = (1, \infty)\), \(A_2 = \{1\}\) and \(A_3 = \{2\}\) and \(C_1=\{1\}, C_2=\{2\}\). If we were to treat instead \(A_1\) as a semi-closed interval \([1, \infty)\), direct maximization of the log likelihood in eq. 2.2 of Turnbull (1976) would give probability half to each observed failure time. By contrast, the Kaplan--Meier estimator, under the convention that right-censored observations at time \(t\) where at risk up to and until \(t\), assigns probability 1/3 to the first failure. To retrieve this solution with Turnbull's EM estimator, we need the convention that \(C_1 \notin A_1\), but this requires comparing the bound with itself. The numerical tolerance in the implementation is taken to be the square root of the machine epsilon for doubles.

The maximum likelihood estimator (MLE) needs not be unique, and the EM algorithm is only guaranteed a local maximum. For interval censored data, Gentleman and Geyer (1994) consider using the Karush--Kuhn--Tucker conditions to determine whether the probability in some intervals is exactly zero and whether the returned value is indeed the MLE.

Due to data scarcity, statistical inference for human lifespan is best conducted using parametric models supported by asymptotic theory, reserving nonparametric estimators to assess goodness-of-fit. The empirical cumulative hazard for Japanese is very close to linear from early ages, suggesting that the hazard may not be very far from exponential even if more complex models are likely to be favored given the large sample size.

The \texttt{longevity} package includes a C\(^{++}\) implementation of the corrected Turnbull's algorithm (Turnbull 1976): the function \texttt{np\_elife} returns a list with Turnbull's intervals \([a_j, b_j]\) and the probability weights assigned to each, provided the latter are positive. It also contains an object of class \texttt{stepfun} with a weighting argument that defines a cumulative distribution function.

\begin{verbatim}
ecdf <- np_elife(arguments = args_japan, thresh = 108)
model2 <- fit_elife(arguments = args_japan,
                    thresh = 108,
                    family = "gp")
plot(ecdf, xlab = "age")
curve(pgpd(q = x,
           scale = coef(model2)['scale'],
           shape = coef(model2)['shape'],
           loc = model2$thresh),
      from = model2$thresh,
      to = model2$thresh + 10,
      add = TRUE)
\end{verbatim}

\begin{figure}

{\centering \includegraphics[width=0.7\linewidth]{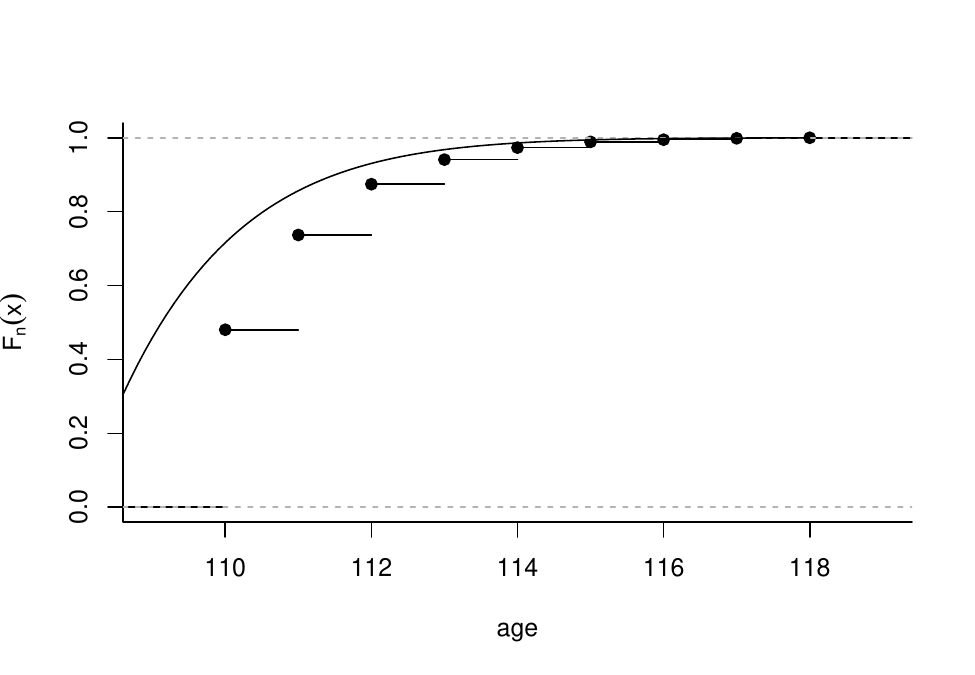}

}

\caption{Nonparametric maximum likelihood estimate of the distribution function with superimposed generalized Pareto fit for excess lifetimes above 108 years.}\label{fig:fig-ecdf}
\end{figure}

\hypertarget{conclusion}{%
\subsection{Conclusion}\label{conclusion}}

This paper describes the salient features of \texttt{longevity}, explaining the theoretical underpinning and the design consideration surrounding the package. While it was conceived for modelling lifetimes in the rare event case, it is generally more applicable. Survival data in extreme value theory is infrequent yet hardly absent and the package could be used for more general analysis: for example, rainfall observations can be viewed as rounded measurements due to their limited precision, with maximum bucket capacity for rain gauge translating into right censoring. While these issues are typically ignored by modelers, \texttt{longevity} offers estimation routines, diagnostic tools and utilities to simplify model assessment. The \texttt{demography} package, which provides forecasting methods for birth and death rates, fertility and population using Lee--Carter and ARIMA models, is complementary to ours.

Some of the features of \texttt{longevity} are interesting in their own right, including the nonparametric maximum likelihood estimator for arbitrary censoring and truncation pattern, as well as adapted quantile-quantile and other visual goodness-of-fit diagnostics for model validation. The testing procedures correctly handle tests for restrictions lying on the boundary of the parameter space. Parametric bootstrap procedures for such tests are not straightforward to implement given the heavy reliance on the data generating mechanism and the diversity of possible scenarios: the paper however shows how the utilities of the package can be coupled to ease such estimation and for testing goodness of fit.

\hypertarget{references}{%
\section*{References}\label{references}}
\addcontentsline{toc}{section}{References}

\hypertarget{refs}{}
\begin{CSLReferences}{1}{0}
\leavevmode\vadjust pre{\hypertarget{ref-Anderson-Bergman:2017}{}}%
Anderson-Bergman, Clifford. 2017a. {``An Efficient Implementation of the {EMICM} Algorithm for the Interval Censored {NPMLE}.''} \emph{Journal of Computational and Graphical Statistics} 26 (2): 463--67. \url{https://doi.org/10.1080/10618600.2016.1208616}.

\leavevmode\vadjust pre{\hypertarget{ref-icensReg}{}}%
---------. 2017b. {``{icenReg}: Regression Models for Interval Censored Data in {R}.''} \emph{Journal of Statistical Software} 81 (12): 1--23. \url{https://doi.org/10.18637/jss.v081.i12}.

\leavevmode\vadjust pre{\hypertarget{ref-Barbi:2018}{}}%
Barbi, Elisabetta, Francesco Lagona, Marco Marsili, James W. Vaupel, and Kenneth W. Wachter. 2018. {``The Plateau of Human Mortality: Demography of Longevity Pioneers.''} \emph{Science} 360 (6396): 1459--61. \url{https://doi.org/10.1126/science.aat3119}.

\leavevmode\vadjust pre{\hypertarget{ref-Beard:1963}{}}%
Beard, Robert E. 1963. {``A Theory of Mortality Based on Actuarial, Biological and Medical Considerations.''} In \emph{Proceedings of the International Population Conference, New York}. Vol. 1. London: International Union for the Scientific Study of Population.

\leavevmode\vadjust pre{\hypertarget{ref-mev-package}{}}%
Belzile, Léo R. et al. 2023. \emph{{mev}: Modelling Extreme Values}. \url{https://CRAN.R-project.org/package=mev}.

\leavevmode\vadjust pre{\hypertarget{ref-ARSIA:2022}{}}%
Belzile, Léo R., Anthony C. Davison, Jutta Gampe, Holger Rootzén, and Dmitrii Zholud. 2022. {``Is There a Cap on Longevity? {A} Statistical Review.''} \emph{Annual Review of Statistics and Its Application} 9: 22--45. \url{https://doi.org/10.1146/annurev-statistics-040120-025426}.

\leavevmode\vadjust pre{\hypertarget{ref-Belzile.Dutang.Northrop.Opitz:2022}{}}%
Belzile, Léo R., Christophe Dutang, Paul J. Northrop, and Thomas Opitz. 2023. {``A Modeler's Guide to Extreme Value Software.''} \emph{Extremes}. Springer. \url{https://doi.org/10.1007/s10687-023-00475-9}.

\leavevmode\vadjust pre{\hypertarget{ref-casebase}{}}%
Bhatnagar, Sahir Rai, Maxime Turgeon, Jesse Islam, James A. Hanley, and Olli Saarela. 2022. {``{casebase}: An Alternative Framework for Survival Analysis and Comparison of Event Rates.''} \emph{The R Journal} 14: 59--79. \url{https://doi.org/10.32614/RJ-2022-052}.

\leavevmode\vadjust pre{\hypertarget{ref-Camarda:2022}{}}%
Camarda, Carlo Giovanni. 2022. {``The Curse of the Plateau. Measuring Confidence in Human Mortality Estimates at Extreme Ages.''} \emph{Theoretical Population Biology} 144: 24--36. \url{https://doi.org/10.1016/j.tpb.2022.01.002}.

\leavevmode\vadjust pre{\hypertarget{ref-Chernoff:1954}{}}%
Chernoff, Herman. 1954. {``On the Distribution of the Likelihood Ratio.''} \emph{The Annals of Mathematical Statistics} 25 (3): 573--78. \url{https://doi.org/10.1214/aoms/1177728725}.

\leavevmode\vadjust pre{\hypertarget{ref-fitdistrplus}{}}%
Delignette-Muller, Marie Laure, and Christophe Dutang. 2015. {``{fitdistrplus}: An {R} Package for Fitting Distributions.''} \emph{Journal of Statistical Software} 64 (4): 1--34. \url{https://doi.org/10.18637/jss.v064.i04}.

\leavevmode\vadjust pre{\hypertarget{ref-Devroye:1986}{}}%
Devroye, L. 1986. \emph{Non-Uniform Random Variate Generation}. New York: Springer. \url{http://www.nrbook.com/devroye/}.

\leavevmode\vadjust pre{\hypertarget{ref-Efron.Petrosian:1999}{}}%
Efron, Bradley, and Vahe Petrosian. 1999. {``Nonparametric Methods for Doubly Truncated Data.''} \emph{Journal of the American Statistical Association} 94 (447): 824--34. \url{https://doi.org/10.1080/01621459.1999.10474187}.

\leavevmode\vadjust pre{\hypertarget{ref-Einmahl:2019}{}}%
Einmahl, Jesson J., John H. J. Einmahl, and Laurens de Haan. 2019. {``Limits to Human Life Span Through Extreme Value Theory.''} \emph{Journal of the American Statistical Association} 114 (527): 1075--80. \url{https://doi.org/10.1080/01621459.2018.1537912}.

\leavevmode\vadjust pre{\hypertarget{ref-interval}{}}%
Fay, Michael P., and Pamela A. Shaw. 2010. {``Exact and Asymptotic Weighted Logrank Tests for Interval Censored Data: The {interval} {R} Package.''} \emph{Journal of Statistical Software} 36 (2): 1--34. \url{https://doi.org/10.18637/jss.v036.i02}.

\leavevmode\vadjust pre{\hypertarget{ref-Frydman:1994}{}}%
Frydman, Halina. 1994. {``A Note on Nonparametric Estimation of the Distribution Function from Interval-Censored and Truncated Observations.''} \emph{Journal of the Royal Statistical Society. Series B (Methodological)} 56 (1): 71--74. \url{https://doi.org/10.1111/j.2517-6161.1994.tb01960.x}.

\leavevmode\vadjust pre{\hypertarget{ref-Gentleman.Geyer:1994}{}}%
Gentleman, Robert, and Charles J. Geyer. 1994. {``Maximum Likelihood for Interval Censored Data: Consistency and Computation.''} \emph{Biometrika} 81 (3): 618--23. \url{https://doi.org/10.1093/biomet/81.3.618}.

\leavevmode\vadjust pre{\hypertarget{ref-Icens-package}{}}%
Gentleman, Robert, and Alain Vandal. 2021. \emph{{Icens}: {NPMLE} for Censored and Truncated Data}.

\leavevmode\vadjust pre{\hypertarget{ref-Rsolnp-pkg}{}}%
Ghalanos, Alexios, and Stefan Theussl. 2015. \emph{{Rsolnp}: General Non-Linear Optimization Using Augmented {L}agrange Multiplier Method}.

\leavevmode\vadjust pre{\hypertarget{ref-extRemes}{}}%
Gilleland, Eric, and Richard W. Katz. 2016. {``{extRemes} 2.0: An Extreme Value Analysis Package in {R}.''} \emph{Journal of Statistical Software} 72 (8): 1--39. \url{https://doi.org/10.18637/jss.v072.i08}.

\leavevmode\vadjust pre{\hypertarget{ref-Groeneboom.Wellner:1992}{}}%
Groeneboom, Piet, and Jon A. Wellner. 1992. \emph{Information Bounds and Nonparametric Maximum Likelihood Estimation}. Edited by Birkhäuser. Basel. \url{https://doi.org/10.1007/978-3-0348-8621-5}.

\leavevmode\vadjust pre{\hypertarget{ref-lubridate-package}{}}%
Grolemund, Garrett, and Hadley Wickham. 2011. {``Dates and Times Made Easy with {lubridate}.''} \emph{Journal of Statistical Software} 40 (3): 1--25. \url{https://www.jstatsoft.org/v40/i03/}.

\leavevmode\vadjust pre{\hypertarget{ref-muhaz-package}{}}%
Hess, Kenneth, and Robert Gentleman. 2021. \emph{{muhaz}: Hazard Function Estimation in Survival Analysis}. \url{https://CRAN.R-project.org/package=muhaz}.

\leavevmode\vadjust pre{\hypertarget{ref-flexsurv}{}}%
Jackson, Christopher. 2016. {``{flexsurv}: A Platform for Parametric Survival Modeling in {R}.''} \emph{Journal of Statistical Software} 70 (8): 1--33. \url{https://doi.org/10.18637/jss.v070.i08}.

\leavevmode\vadjust pre{\hypertarget{ref-IDL:2021}{}}%
Jdanov, Dmitri A., Vladimir M. Shkolnikov, and Sigrid Gellers-Barkmann. 2021. {``{The International Database on Longevity: Data Resource Profile}.''} In \emph{Exceptional Lifespans}, edited by Heiner Maier, Bernard Jeune, and James W. Vaupel, 22--24. Demographic Research Monographs. Cham, Switzerland: Springer.

\leavevmode\vadjust pre{\hypertarget{ref-Kaplan.Meier:1958}{}}%
Kaplan, Edward L., and Paul Meier. 1958. {``Nonparametric Estimation from Incomplete Observations.''} \emph{Journal of the American Statistical Association} 53: 457--81. \url{https://doi.org/10.1080/01621459.1958.10501452}.

\leavevmode\vadjust pre{\hypertarget{ref-Lindsey.Ryan:1998}{}}%
Lindsey, Jane C., and Louise M. Ryan. 1998. {``Methods for Interval-Censored Data.''} \emph{Statistics in Medicine} 17 (2): 219--38. \url{https://doi.org/10.1002/(SICI)1097-0258(19980130)17:2\%3C219::AID-SIM735\%3E3.0.CO;2-O}.

\leavevmode\vadjust pre{\hypertarget{ref-rstpm2}{}}%
Liu, Xing-Rong, Yudi Pawitan, and Mark Clements. 2018. {``Parametric and Penalized Generalized Survival Models.''} \emph{Statistical Methods in Medical Research} 27 (5): 1531--46. \url{https://doi.org/10.1177/0962280216664760}.

\leavevmode\vadjust pre{\hypertarget{ref-ExceptionalLifespans}{}}%
Maier, Heiner, Bernard Jeune, and James W. Vaupel, eds. 2021. \emph{Exceptional Lifespans}. Demographic Research Monographs. Cham, Switzerland: Springer. \url{https://doi.org/10.1007/978-3-030-49970-9}.

\leavevmode\vadjust pre{\hypertarget{ref-DTDA-package}{}}%
Moreira, Carla, Jacobo de Uña-Álvarez, and Rosa Crujeiras. 2022. \emph{DTDA: Doubly Truncated Data Analysis}. \url{https://CRAN.R-project.org/package=DTDA}.

\leavevmode\vadjust pre{\hypertarget{ref-Northrop.Coleman:2014}{}}%
Northrop, Paul J., and Claire L. Coleman. 2014. {``Improved Diagnostic Plots for Extreme Value Analyses.''} \emph{Extremes} 17: 289--303. \url{https://doi.org/10.1007/s10687-014-0183-z}.

\leavevmode\vadjust pre{\hypertarget{ref-Perks:1932}{}}%
Perks, Wilfred. 1932. {``On Some Experiments in the Graduation of Mortality Statistics.''} \emph{Journal of the Institute of Actuaries} 63 (1): 12--57. \url{https://doi.org/10.1017/S0020268100046680}.

\leavevmode\vadjust pre{\hypertarget{ref-bshazard}{}}%
Rebora, Paola, Agus Salim, and Marie Reilly. 2014. {``{bshazard}: A Flexible Tool for Nonparametric Smoothing of the Hazard Function.''} \emph{{The R Journal}} 6 (2): 114--22. \url{https://doi.org/10.32614/RJ-2014-028}.

\leavevmode\vadjust pre{\hypertarget{ref-Richards:2012}{}}%
Richards, Stephens J. 2012. {``A Handbook of Parametric Survival Models for Actuarial Use.''} \emph{Scandinavian Actuarial Journal} 2012 (4): 233--57. \url{https://doi.org/10.1080/03461238.2010.506688}.

\leavevmode\vadjust pre{\hypertarget{ref-Rotnitzky:2000}{}}%
Rotnitzky, Andrea, David R. Cox, Matteo Bottai, and James Robins. 2000. {``{Likelihood-based inference with singular information matrix}.''} \emph{Bernoulli} 6 (2): 243--84. \url{https://doi.org/bj/1081788028}.

\leavevmode\vadjust pre{\hypertarget{ref-Royston.Parmar:2002}{}}%
Royston, Patrick, and Mahesh K. B. Parmar. 2002. {``Flexible Parametric Proportional-Hazards and Proportional-Odds Models for Censored Survival Data, with Application to Prognostic Modelling and Estimation of Treatment Effects.''} \emph{Statistics in Medicine} 21 (15): 2175--97. \url{https://doi.org/10.1002/sim.1203}.

\leavevmode\vadjust pre{\hypertarget{ref-Sailynoja.Burkner.Vehtari:2021}{}}%
Säilynoja, Teemu, Paul-Christian Bürkner, and Aki Vehtari. 2022. {``Graphical Test for Discrete Uniformity and Its Applications in Goodness of Fit Evaluation and Multiple Sample Comparison.''} \emph{Statistics and Computing} 32 (2): 32. \url{https://doi.org/10.1007/s11222-022-10090-6}.

\leavevmode\vadjust pre{\hypertarget{ref-Self.Liang:1987}{}}%
Self, Steven G., and Kung-Yee Liang. 1987. {``Asymptotic Properties of Maximum Likelihood Estimators and Likelihood Ratio Tests Under Nonstandard Conditions.''} \emph{Journal of the American Statistical Association} 82 (398): 605--10. \url{https://doi.org/10.1080/01621459.1987.10478472}.

\leavevmode\vadjust pre{\hypertarget{ref-Shen:2010}{}}%
Shen, Pao-Sheng. 2010. {``Nonparametric Analysis of Doubly Truncated Data.''} \emph{Annals of the Institute of Statistical Mathematics} 62 (5): 835--53. \url{https://doi.org/10.1007/s10463-008-0192-2}.

\leavevmode\vadjust pre{\hypertarget{ref-evd}{}}%
Stephenson, A. G. 2002. {``{evd}: Extreme Value Distributions.''} \emph{R News} 2 (2). \url{https://CRAN.R-project.org/doc/Rnews/}.

\leavevmode\vadjust pre{\hypertarget{ref-Sun.Genton:2011}{}}%
Sun, Ying, and Marc G. Genton. 2011. {``Functional Boxplots.''} \emph{Journal of Computational and Graphical Statistics} 20 (2): 316--34. \url{https://doi.org/10.1198/jcgs.2011.09224}.

\leavevmode\vadjust pre{\hypertarget{ref-survival-package}{}}%
Therneau, Terry M. 2022. \emph{A Package for Survival Analysis in {R}}. \url{https://CRAN.R-project.org/package=survival}.

\leavevmode\vadjust pre{\hypertarget{ref-survival-book}{}}%
Therneau, Terry M., and Patricia M. Grambsch. 2000. \emph{Modeling Survival Data: Extending the {C}ox Model}. New York: Springer. \url{https://doi.org/10.1007/978-1-4757-3294-8}.

\leavevmode\vadjust pre{\hypertarget{ref-Tsai.Jewell.Wang:1987}{}}%
Tsai, Wei-Yann, Nicholas P. Jewell, and Mei-Cheng Wang. 1987. {``A Note on the Product-Limit Estimator Under Right Censoring and Left Truncation.''} \emph{Biometrika} 74 (4): 883--86. \url{https://doi.org/10.1093/biomet/74.4.883}.

\leavevmode\vadjust pre{\hypertarget{ref-Turnbull:1976}{}}%
Turnbull, Bruce W. 1976. {``The Empirical Distribution Function with Arbitrarily Grouped, Censored and Truncated Data.''} \emph{Journal of the Royal Statistical Society, Series B} 38: 290--95. \url{https://doi.org/10.1111/j.2517-6161.1976.tb01597.x}.

\leavevmode\vadjust pre{\hypertarget{ref-Waller.Turnbull:1992}{}}%
Waller, Lance A., and Bruce W. Turnbull. 1992. {``Probability Plotting with Censored Data.''} \emph{American Statistician} 46 (1): 5--12. \url{https://doi.org/10.1080/00031305.1992.10475837}.

\leavevmode\vadjust pre{\hypertarget{ref-ggplot2}{}}%
Wickham, Hadley. 2016. \emph{{ggplot2}: Elegant Graphics for Data Analysis}. Springer-Verlag New York. \url{https://ggplot2.tidyverse.org}.

\leavevmode\vadjust pre{\hypertarget{ref-dplyr-package}{}}%
Wickham, Hadley, Romain François, Lionel Henry, Kirill Müller, and Davis Vaughan. 2023. \emph{{dplyr}: A Grammar of Data Manipulation}. \url{https://CRAN.R-project.org/package=dplyr}.

\leavevmode\vadjust pre{\hypertarget{ref-Ye:1987}{}}%
Ye, Yinyu. 1987. {``Interior Algorithms for Linear, Quadratic, and Linearly Constrained Non-Linear Programming.''} PhD thesis, Department of {ESS}, Stanford University.

\leavevmode\vadjust pre{\hypertarget{ref-VGAMbook}{}}%
Yee, Thomas W. 2015. \emph{Vector Generalized Linear and Additive Models: With an Implementation in {R}}. New York, NY: springer. \url{https://doi.org/10.1007/978-1-4939-2818-7}.

\leavevmode\vadjust pre{\hypertarget{ref-Yee.Wild:1996}{}}%
Yee, Thomas W., and C. J. Wild. 1996. {``Vector Generalized Additive Models.''} \emph{Journal of the Royal Statistical Society: Series B (Methodological)} 58 (3): 481--93. \url{https://doi.org/10.1111/j.2517-6161.1996.tb02095.x}.

\leavevmode\vadjust pre{\hypertarget{ref-evgam}{}}%
Youngman, Benjamin D. 2022. {``{evgam}: An {R} Package for Generalized Additive Extreme Value Models.''} \emph{Journal of Statistical Software} 103 (3): 1--26. \url{https://doi.org/10.18637/jss.v103.i03}.

\end{CSLReferences}

\end{document}